\documentclass[final,3p,sort&compress,times,single,english]{elsarticle}
\usepackage{lineno,hyperref}
\usepackage{amsmath}
\usepackage{amsthm}
\usepackage{amsfonts}
\usepackage{natbib}
\usepackage{lipsum}
\usepackage[bf]{caption}
\usepackage{caption}
\usepackage[T1]{fontenc}
\usepackage[utf8]{inputenc}
\usepackage{babel}
\usepackage[export]{adjustbox}
\usepackage{array}
\usepackage{setspace}
\usepackage{geometry}
\usepackage{array}
\usepackage{tikz}
\usepackage{xcolor}
\usetikzlibrary{shapes,shapes.geometric,arrows,fit,calc,positioning,automata,}
\usetikzlibrary{automata,positioning}
\usepackage{varwidth}
\usepackage{cleveref}
\usepackage{mathrsfs}
\usepackage{epstopdf}
\usepackage[toc,page]{appendix}

\usepackage{algorithm}
\usepackage{multicol}
\usepackage{arydshln}
\renewcommand{\thefootnote}{\emph{\alph{footnote}}}
\usepackage{fancyhdr}
\usepackage{caption}
\hypersetup{colorlinks=true,linkcolor=green,filecolor=blue,urlcolor=blue}
\makeatletter
\def\ps@pprintTitle{%
	\let\@oddhead\@empty
	\let\@evenhead\@empty
	\def\@oddfoot{\normalfont
 \ifx\@journal
		\else\@journal\fi\hfil\thepage\hspace{6.45in}\thepage}
\let\@evenfoot\@oddfoot}

\renewcommand\@biblabel[1]{#1} 
\makeatother

\begin{document}
\begin{frontmatter}
\title{Statistical Learning in Computed Tomography Image Estimation}
\author[1]{Fekadu L. Bayisa}
\author[1]{Xijia Liu}
\author[2]{Anders Garpebring}
\author[1]{and Jun Yu}
\address[1]{Department of Mathematics and Mathematical Statistics, Ume{\aa} University, Ume{\aa}, 901 87, Sweden}
\address[2]{Department of Radiation Sciences, Ume{\aa} University, Ume{\aa}, 901 87, Sweden}
\cortext[author] {Corresponding author.\\\textit{E-mail address:} fekadu.bayisa@umu.se}
\begin{abstract}
\textbf{Purpose}: There is increasing interest in computed tomography (CT) image estimations from magnetic resonance (MR) images. The estimated CT images can be utilised for attenuation correction, patient positioning, and dose planning in diagnostic and radiotherapy workflows. This study aims to introduce a novel statistical learning approach for improving CT estimation from MR images and to compare the performance of our method with the existing model based CT image estimation methods.\\
\textbf{Methods}: The statistical learning approach proposed here consists of two stages. At the training stage, prior knowledges about tissue-types from CT images were used together with a Gaussian mixture model (GMM) to explore CT image estimations from MR images. Since the prior knowledges are not available at the prediction stage, a classifier based on RUSBoost algorithm was trained to estimate the tissue-types from MR images. For a new patient, the trained classifier and GMMs were used to predict CT image from MR images. The classifier and GMMs were validated by using voxel-level 10-fold cross-validation and patient-level leave-one-out cross-validation, respectively. \\ 
\textbf{Results}: The proposed approach has outperformance in CT estimation quality in comparison with the existing model-based methods, especially on bone tissues. Our method improved CT image estimation by $5\%$ and $23\%$ on the whole brain and bone tissues, respectively. \\
\textbf{Conclusions}: Evaluation of our method shows that it is a promising method to generate  CT image substitutes for the implementation of fully MR-based radiotherapy and PET/MRI applications.

\end{abstract}
\end{frontmatter}

\renewcommand{\thefootnote}{\emph{\alph{footnote}}}
\section{Introduction}
Diagnostic imaging technologies such as magnetic resonance imaging (MRI) and computed tomography (CT) are used to obtain detailed anatomical images, although the ionising radiation of CT is associated with an increased risk of radiation-induced cancer  \citep{Mathews}. MRI is suitable for tumour localisation, which increases the effectiveness of radiation dose planning \cite{SuitH, Just, Heesters}. The interest in removing MRI-to-CT image fusion uncertainties, developing MRI linear accelerators, and increasing the tumour-localising ability of MRI have encouraged the development of fully MRI-based radiotherapy planning \cite{BoydevO}. However, it is challenging for MRI to delineate solid bone structures due to the short signal lifetimes and low proton densities in bone tissue \cite{WiesingerF}. Besides, MR images do not provide electron density information for dose calculation. In addition to its geometrical and intensity distortions, MRI also lacks digitally reconstructed radiograph generation for verification of patient positioning  \cite{Chengu,Karger, BoydevO}.  Therefore, coregistered CT and MR images have been regarded as a complementary procedure \cite{Boettger, Karlssonr}.

Recently, there has been a growing interest in the development and application of integrated positron emission tomography (PET)/MRI scanners, which can acquire PET and MR images simultaneously in brain imaging studies \cite{HerzogH, CatanaC, CatanaC2}.
The implementation of integrated PET/MRI scanners requires information about electron densities for attenuation correction \cite{BurgosNand}, and this information is obtained from CT images. Although MR images can be obtained from the integrated scanner, they do not provide attenuation related measurements of electron densities to develop a reliable CT image estimation method for the integration of PET/MRI scanners.

The estimated CT images can be used to enable accurate MRI-based radiotherapy, and they are essential for attenuation correction in PET imaging.  In 2013,  Hsu et al. \cite{HsuSHJ2} utilised probabilistic classification of voxels to obtain CT images that are currently supporting the workflows of radiation oncology treatment planning in the brain. Furthermore, solutions to the problem of CT substitute generation are becoming available commercially \cite{SiverssonC23}. Several CT image estimation methods have been reported, and the estimated CT images are acceptable for clinical implementation. However, the lack of established approach for evaluating the performance and reporting the consistency of these estimation methods has challenged and delayed widespread clinical implementation \cite{EdmundNyholm}.

In 2016, Huynh et al. \cite{Huynh} used a structured random forest and auto-context model to estimate CT images from MR images, but this approach is subjective and its performance depends on the quality of features that are extracted from the MR images. A three-dimensional fully convolutional neural network model has also been developed for patch-wise estimation of CT images from MR images \cite{Nie}. In terms of mean absolute error and peak signal-to-noise ratio, this method has better performance on experimental data than structured random forest and atlas-based methods. However, its success depends on the geometry of the data. In 2016, Arabi et al. proposed a two-step atlas-based algorithm to obtain estimated CT images from MR image sequences \cite{Arabi}. Their method provides better bone tissue identification accuracy than the conventional segmentation and atlas-based methods. A Gaussian mixture model (GMM) has also been utilised to explore CT image estimation and the associated estimation  uncertainty \cite{Johansson2, Adam}. Taking contextual information into account, Kuljus et al. \cite{Kuljus} have investigated CT image estimation using a hidden Markov model (HMM) and a Markov random field model (MRF). In terms of mean absolute error, HMM has better performance than MRF and GMM. In comparison with HMM and GMM, MRF has superior estimation quality in bone tissues. However, MRF is computationally expensive. Tissue-dependent performance of the statistical models is the main challenge in CT image estimation, where the estimation quality is found to be poor on air and bone tissues \cite{Adam, Kuljus}. This challenge led Bayisa and Yu \cite{FekaduLB} to examine CT image estimation by partitioning the training data into bone and non-bone tissues, and they used GMM and a skewed-Gaussian mixture model (SGMM) to estimate CT image intensities on each partition. Compared to HMM and MRF, the partitioning approach has better estimation on dense bone tissues. The predictive quality of the partitioning approach depends on the performance of GMM and SGMM on the full data, which is the data without partitioning. However, it is evident that GMM has poor CT image estimation quality on the full data \cite{Kuljus}.

Most of the statistical models that have been used to estimate CT images assume that the images depict different tissue types, which were partly confirmed by the dependence between the performance of models and tissue types. The poor performance on air and bone tissues is the main challenge in statistical models for CT image estimations. This issue has motivated the present work to further investigate and improve the quality of CT image estimation by combining a supervised statistical learning method and GMM. To handle the aforementioned challenge,  we have introduced prior knowledges about the tissue types  into a statistical model and  proposed an algorithm to overcome the class imbalance problem that can arise due to the prior knowledge incorporated to the statistical model. The novelty of our approach is that the prior knowledges about the tissue types are used by the algorithm in such a way that the minority class, which is the bone tissues, is discernible during CT image estimation process.

The main purpose of this article is to further explore CT image estimation and to improve the quality of CT image prediction using statistical learning methods. To achieve this, we propose a classification method, the RUSBoost algorithm, at a supervised learning stage of the CT image estimation process. The RUSBoost algorithm uses random undersampling in an adaptive boosting algorithm to overcome the class imbalance problem in the classification of real datasets \cite{GalarM, SeiffertC}. Decision trees are exploited in the RUSBoost algorithm as weak classifiers, which are classifiers that perform slightly better than random guessing, because they are usually taken to be the best out-of-the-box learners \cite{PraveenamM}. The trained classifier is then used to predict the tissue types of the new MR images from which we estimate the corresponding CT image. The predicted tissue types and GMM are utilised to obtain the desired estimated CT image. Hereafter we call our method RGMM. We have also compared the CT image estimation performance of RGMM, HMM, GMM, MRF, and GMM$^{*}$ (when GMM is used at the supervised learning stage, see \cite{FekaduLB}).

This article is organised as follows. Section \ref{sec2} describes the details of our approach, and the results of the study are presented in Section \ref{sect3}.  We discuss the implication of the results in Section \ref{sec4} and present the conclusions of the study in Section \ref{sec5}.
\section{Materials and Methods} \label{sec2}
In this section, we first describe data acquisition and image registration, then explain the details of the proposed method. Regarding to the method part, we present GMMs which capture the distribution of data given the prior knowledges, and explain how the RUSBoost algorithm trains a classifier to estimate the tissue types. The classifier and GMMs are then combined to make the prediction feasible. At the end, model validation methods are presented. 
\subsection{Data acquisition}
\begin{figure}[H]
    \centering
	\includegraphics[width=80mm, height=80mm]{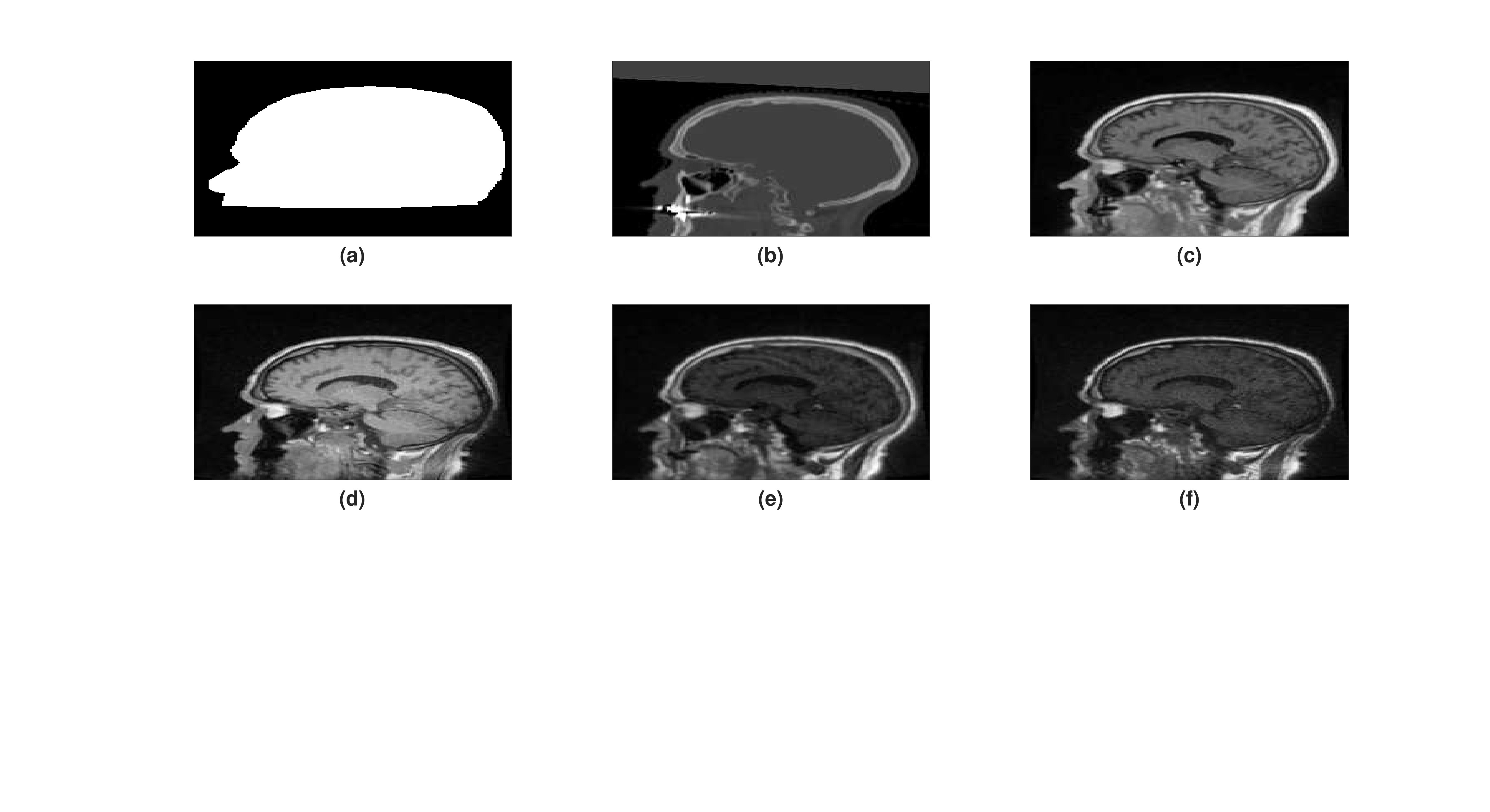}
	\vspace{-1.2in}
	\small\caption{ Binary mask (a), CT image (b) and MR images (c-f).}
	\label{Or1}
\end{figure}

Three-dimensional CT and MR images were acquired from the heads of nine patients. Four MR images were obtained from each patient using two dual echo UTE sequences with flip angles of 10 and 30 degrees. The UTE sequences were sampled from a first echo (free induction decay) and a second echo (gradient echo) from the same excitation with echo times of 0.07 and 3.76 milliseconds, respectively. Each MR image was reconstructed to a  192 $\times$ 192 $\times$ 192 dimensional matrix.  An entry in the matrix represents a signal intensity corresponding to a voxel with size  1.33 $\times$ 1.33 $\times$ 1.33 mm$^{3}$. For each patient, one CT image was acquired using a GE Lightspeed with a 2.5 mm slice thickness, and the image was reconstructed with an in-plane resolution of  0.78  $\times$ 0.78 mm$^{2}$. A binary mask was also developed for each patient to demarcate the area of interest. The binary mask, the CT image, and the four MR images of each patient were co-registered and resampled to the same resolution using linear interpolation. The MR images were pre-processed and normalized, and for further technical details, we refer to \cite{Johansson2, johansson2014magnetic}. We organized the voxel values of the CT image, the binary mask, and the four MR images into six columns to obtain data for each patient. The organised data of the patients were column stacked, and data corresponding to the region of interest were selected for model fitting. 
Fig. \ref{Or1} shows a sagittal image slice for a representative patient.
\subsection{Image registration}
CT image and MR images of each patient were coregistered using a mutual information algorithm from the Insight Toolkit \citep{yoo2002medicine}. A manual evaluation of the registration accuracy has been implemented and each registration passed the evaluation. Using linear interpolation, we resampled the CT image  to have the same resolution as the UTE images.
\subsection{Gaussian mixture model}
Let $Y_{i}$ and $\mathbf{X}_{i} = \left(X_{i1}, {X}_{i2}, \cdots, {X}_{id} \right) $ represent the CT image intensity and its corresponding d-dimensional intensities of the MR images at voxel  $i$. In our real data, we have four MR images, $d=4$.  Let $t_{i}$ be its corresponding known tissue type, which represents prior knowledge about tissue types. The class probability density of the random vector $\left(Y_{i}, \mathbf{X}_{i}\right)$ is assumed to be Gaussian:

\begin{equation}
\left(Y_{i}, \mathbf{X}_{i}\right) \mid t_{i} = k \sim \mathcal{N}(\boldsymbol\mu_{k}, \boldsymbol\Sigma_{k}),
\end{equation}
where $i = 1, 2, \cdots, n,$  $n$ is the number of voxels, $\boldsymbol\mu_{k}$ is the mean vector, $\boldsymbol\Sigma_{k}$ is the variance-covariance structure, $k = 1, 2, \cdots, K,$  and $K$ is the known number of classes. In this study, $K = 2$. For each known tissue type, we assume that there is a latent variable $Z_{i}$ that represents an unobserved subtissue type at voxel $i$. To take this into account, we have assumed a more flexible model given by

\begin{equation}
\left(Y_{i}, \mathbf{X}_{i}\right) \mid t_{i} = k, Z_{i}=j \sim \mathcal{N}(\boldsymbol\mu_{kj}, \boldsymbol\Sigma_{kj}),
\end{equation}
where $j=1,  2, \cdots, J_{k},$ and $J_{k}$ is the number of subtissue types within the known tissue type $k$. Consequently, we have a Gaussian mixture model (GMM) for each known tissue type, and this is given by
\begin{equation}
\left(Y_{i}, \mathbf{X}_{i}\right)\mid t_{i} = k\sim \sum_{j=1}^{J_{k}}\pi_{kj}\mathcal{N}(\boldsymbol\mu_{kj}, \boldsymbol\Sigma_{kj}),
\end{equation}
where
 \begin{eqnarray}	
	\pi_{kj} \ge 0,\quad \sum_{j=1}^{J_{k}}\pi_{kj}=1,\quad k= 1, 2, \cdots, K.
\end{eqnarray}

The GMM parameters can be estimated by an iterative maximising procedure using the concept of incomplete data via the EM-algorithm  \cite{Dempster}. We use the mean squared error to select the optimal estimates.

Two stages are used during the parameter estimation process. First, we estimate the model parameters given the number of unknown tissue types. The mean squared error is utilised to select the optimal parameter estimates for the given number of unknown tissue types. Second, we select the number of unknown tissue types by using the estimated models on the validation dataset. The final optimal parameter estimates are the ones that have smaller mean squared errors or mean absolute errors for the validation dataset. The stopping criterion used by Bayisa and Yu. \cite{FekaduLB} is applied to control the convergence of the estimation process.
\subsection{Supervised learning method}
Using the observed CT image intensity $y_{i}$ that corresponds to the intensity $\mathbf{x}_{i}$ of the MR images, we define a class label or tissue type of the MR images at voxel $i$ as follows.

\begin{equation}
t_{i}  =
\begin{cases}
0, & \text{if} \hspace{0.1in}y_{i}\leq 100\hspace{0.03in}\text{HU}, \\
1, & \text{otherwise},
\end{cases}
\end{equation}
where $100$ HU is used as CT image intensity threshold value (see \cite{FekaduLB}) and HU is a Hounsfield unit.
\subsubsection{RUSBoost algorithm}
Important information in datasets can be captured by well-conceived new features. It is reasonable to consider spatial features during the CT image estimation process. CT and MR images are three-dimensional images, and a given voxel in the images that is not on the boundary of the images has six first-order and a maximum of twenty-six second-order neighbouring voxels.  Let $\mathbf{x}^{s}_{i}$ be a vector of intensities of MR images at the six or the twenty-six closest voxels to voxel $i$,  $\mathbf{x}^{c}_{i} = \left(\mathbf{x}_{i}, \mathbf{x}^{s}_{i} \right)$ be a vector of the original and the extracted spatial features at voxel $i$, and $t_{i}$ be its corresponding known class label or tissue type. Based on $x_i^c$ and $t_i$, the RUSBoost algorithm is used to train a supervised learning method. Decision trees have been adopted as weak learners in the RUSBoost algorithm, and we have explored all possible binary splits on every predictor. The optimisation of the binary split is subjected to the maximum number of splits, the purity, and the minimum leaf size of a node. The split predictor among all possible splits of all predictors can be obtained using the GINI index given by
\begin{eqnarray}
	GINI(v) = 1-\sum_{t=1}^{K}p^2_{tv},
\end{eqnarray}
where $K$ is the number of class labels, $v$ is a node, and $p_{tv}$ is the proportion of the class label $t$ at node $v$. GINI index is a measure of node impurity and minimizing it can lead to the purity of the node.
The RUSBoost algorithm is shown in Algorithm \ref{Oromia202}. The algorithm can be explained as follows. For each  iteration  $j$ = 1, 2, $\cdots$, $M$, the RUSBoost algorithm calls for an external weak learning algorithm and provides it with the data and the mislabel distribution. The weak learning algorithm returns a weak learner, which is a function $h_{j}$: ${\Bbb X} \times {\Bbb T} \rightarrow \left[0, 1\right]$, to the RUSBoost algorithm. One can give meaning to a weak learner $h_{j}$ in the following manner. If $h_{j}\left( \mathbf{x}^{c}_{i}, t_{i}\right) = 1$ and $h_{j}\left( \mathbf{x}^{c}_{i}, t\right) = 0$ then $h_{j}$ has correctly identified the label at voxel ${i}$ as $t_{i}$. In a similar manner, if $h_{j}\left( \mathbf{x}^{c}_{i}, t_{i}\right) = 0$  and $h_{j}\left( \mathbf{x}^{c}_{i}, t\right) = 1$, then $h_{j}$  has incorrectly identified the label at voxel ${i}$ as $t$. If $h_{j}\left(\mathbf{x}^{c}_{i}, t_{i}\right)$ =  $h_{j}\left( \mathbf{x}^{c}_{i}, t\right)$, then  label prediction by $h_{j}$ is taken to be a random guess. Even though the function $h_{j}$ may not define a proper probability distribution, the values of $h_{j}\in\left(0, 1 \right)$ can be interpreted probabilistically \citep{FreundYS}.

The weak learners are trained sequentially. The first weak learner is trained on a set of instances obtained by random undersampling (RUS) from a training set.  The training set for the second weak learner is obtained by RUS from instances that are picked with replacement from the training set with higher probabilities, which is according to the mislabel distribution assigned to those observations that are classified incorrectly by the first weak learner.  Instances that are difficult to classify are more likely to occur multiple times in the training set because they are picked with replacement. As we proceed to build each weak learner in the ensemble, instances which are more difficult to classify correctly appear more and more likely. Since each weak learner is asked to classify the more difficult instances, the training error of an individual weak learner tends to increase as the number of weak learners increases. However, the boosting algorithm shows that the ensemble training error rate decreases as the number of weak learners increases. The ensemble output is determined by weighting the weak learner with $\log(1/\alpha_{j})$, where $\alpha_{j}$ is proportional to the $j$th weak learner error rate. If the weak learner has good error rate performance, it  has good contribution to the output. At a given voxel $i$, the final classifier $H$ predicts a label $v\in {\Bbb T}$ that maximizes a weighted average of the weak learners.

\begin{algorithm}[H]
	\caption{\textbf{RUSBoost} Algorithm}
	\begin{enumerate}[a)]\label{Oromia202}
		\item Let  $\mathbf{D}$ be a training set given by  	
		\begin{equation*}
		\mathbf{D}=\left\lbrace \left(\mathbf{x}^{c}_{1}, t_{1}\right), \left(\mathbf{x}^{c}_{2}, t_{2}\right), \cdots, \left(\mathbf{x}^{c}_{n}, t_{n}\right)\right\rbrace,
		\end{equation*}
		where  $\mathbf{x}^{c}_{i} \in {\Bbb X}$, ${t}_{i} \in {\Bbb T}$, $i= 1, 2, \cdots, n$, ${\Bbb X}$ is a feature space and ${\Bbb T}$ is a label set;
		\item Random undersampling (RUS) based on a minority class label in ${\Bbb T}$;
		\item Maximum number $M$ of learners in the ensemble;
		\item Weak learning algorithm \textbf{WeakLearn};
		\item Let  $\mathbf{S}=\left\lbrace\left(i, t\right)\mid i = 1, 2, \cdots, n,  t \neq  t_{i}\right\rbrace$ be the set of possible mislabels. In this notation, $i$ and $t$ represent the index of $\mathbf{x}^{c}_{i}$  and one of the incorrect labels of $\mathbf{x}^{c}_{i}$;
		\item Using $\mathbf{S}$, initialise the mislabel distribution $\mathbf{W}_{1}$, which is given by
		\begin{equation*}
		\mathbf{W}_{1} = \left\lbrace W_{1}\left(i, t\right)=\frac{1}{|\mathbf{S}|}: \left(i, t\right)\in\mathbf{S}\right\rbrace,
		\end{equation*}
		where $|\mathbf{S}|$ denotes the cardinality of the set $\mathbf{S}$;
		\item For $j$ = 1, 2, $\cdots$, $M$, do 
		\begin{enumerate}[1.]
			\item Using RUS, obtain the modified training dataset $\mathbf{D}'_{j}$ and its distribution $\mathbf{W}'_{j}$;\label{ORAwaday1}
			\item Call \textbf{WeakLearn} and provide it with the training dataset $\mathbf{D}'_{j}$ and $\mathbf{W}'_{j}$;\label{ORAwaday2}
			\item Obtain a weak classifier $h_{j}$: ${\Bbb X} \times {\Bbb T} \rightarrow \left[0, 1\right]$;
			\item Compute the pseudo-loss of the classifier $h_{j}$
			\begin{equation*}\label{ORAwaday3}
			\epsilon_{j}=\displaystyle\sum_{\left(i, t\right)\in\mathbf{S}}W_{j}\left(i, t\right)\left[1-h_{j}\left(\mathbf{x}^{c}_{i}, t_{i}\right) + h_{j}\left(\mathbf{x}^{c}_{i}, t\right)\right] ;
			\end{equation*}
			\item Compute the weight update parameter $\alpha_{j}$
			\begin{equation*}
			\alpha_{j}=\frac{\epsilon_{j}}{1-\epsilon_{j}};
			\end{equation*}
			\item Update the weight $W_{j}\left(i, t\right)$
			\begin{equation*}
			W_{j+1}\left(i, t\right)=W_{j}\left(i, t\right)\alpha_{j}^{\frac{1}{2}\left[1+h_{j}\left  (\mathbf{x}^{c}_{i}, t_{i}\right) - h_{j}\left(\mathbf{x}^{c}_{i}, t \right)\right]};
			\end{equation*}
			\item Normalise the weight $ W_{j+1}\left(i, t\right)$
			\begin{equation*}
			W_{j+1}\left(i, t\right)=\frac{W_{j+1}\left(i, t\right)}{\displaystyle\sum_{\left(i, t\right)\in \mathbf{S}}W_{j+1}\left(i, t\right)};
			\end{equation*}
		\end{enumerate}
		\item Obtain the final strong learner
		\begin{equation*}
		H\left(\mathbf{x}^{c}_{i}\right)=\arg\hspace{-0.03in}\max\limits_{v\in {\Bbb T}} \displaystyle\sum_{j=1}^{M}h_{j}\left  (\mathbf{x}^{c}_{i}, v\right)\log\frac{1}{\alpha_{j}};
		\end{equation*}
	\end{enumerate}
\end{algorithm}

\subsection{Estimation of CT images}
We are interested in estimating the CT image intensity $Y_{i}$ given new MR images $\mathbf{X}_{i}$ and its known tissue type $t_{i}$ at voxel $i$. A point estimator of CT image intensity can be given by
	\begin{equation}\label{expectedvalue} 
	E\left[Y_{i}\mid \mathbf{X}_{i}, t_{i}=k, \boldsymbol{\Theta}\right] 
	=\sum_{j=1}^{J_{k}}\beta_{j} E\left[Y_{i}\mid \mathbf{X}_{i}, t_{i}=k, Z_{i} =j, \boldsymbol{\Theta}\right],
	\end{equation}
where 
\begin{equation}
\boldsymbol\Theta=\left\lbrace\pi_{kj}, \boldsymbol\mu_{kj}, \boldsymbol\Sigma_{kj}\mid j= 1, 2, \cdots, J_{k};  k= 1, 2, \cdots, K\right\rbrace,
\end{equation}
is the set of all parameters. One can compute the weight $\beta_{j}=P\left(Z_{i} =j\mid \mathbf{X}_{i}, t_{i}=k, \boldsymbol{\Theta}\right)$ by Bayes' theorem. The expected value  $E\left[Y_{i}\mid \mathbf{X}_{i}, t_{i}=k, Z_{i} =j, \boldsymbol{\Theta}\right]$ is the conditional expectation in the multivariate normal distribution. The first-layer tissue type  $t_{i}$ for the MR images at voxel $i$ can be estimated using the RUSBoost algorithm at a supervised learning stage. The predicted tissue types  and the estimated GMM parameters are exploited to obtain the desired CT images from the new MR images. 

\begin{figure}[H]
		\centering
		\includegraphics[width=80mm, height=60mm]{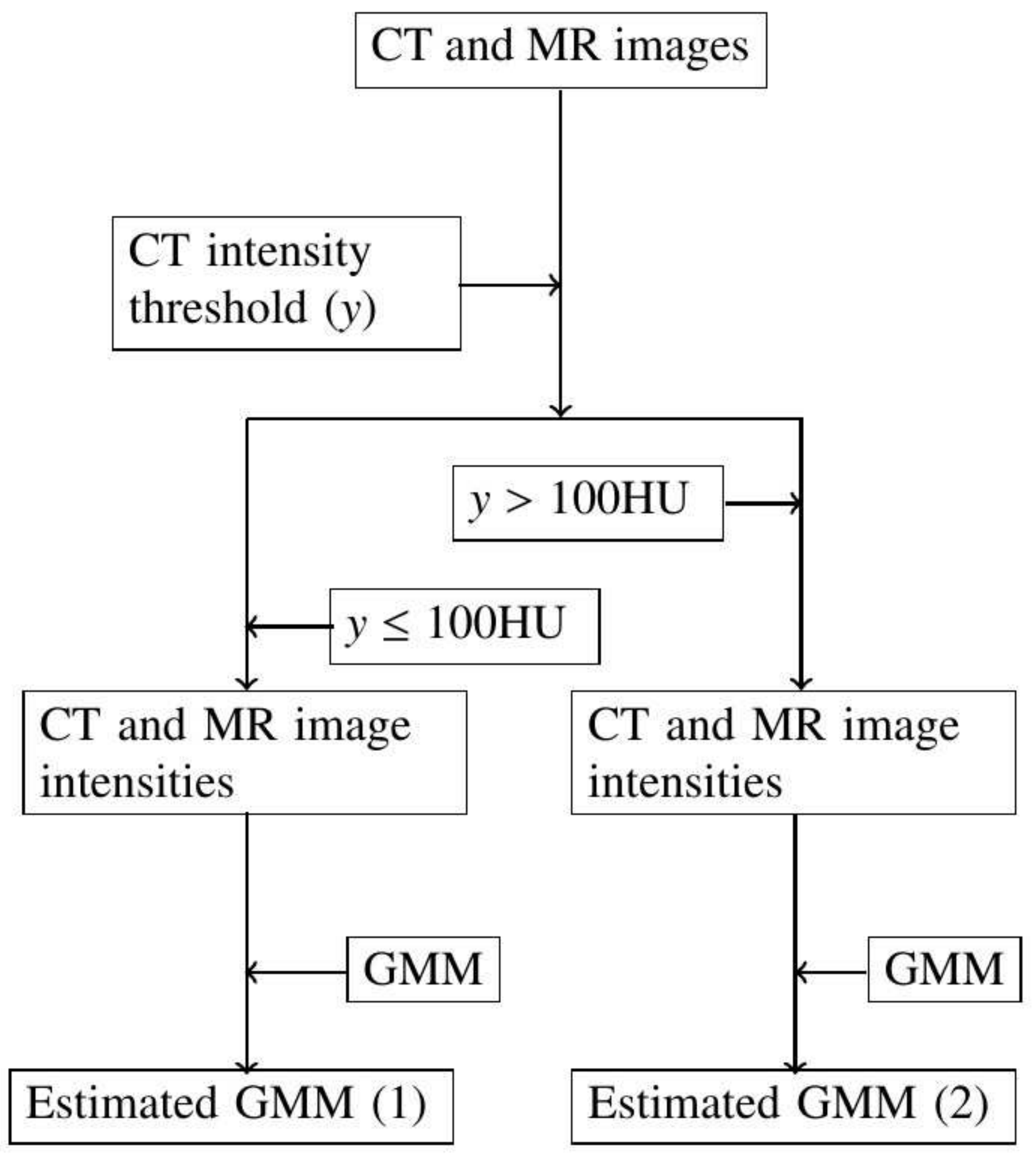}
		\caption{Conceptual model for training GMM.}
		\label{fig:Pictorialed}
\end{figure}

\begin{figure}[H]
		\centering
	\includegraphics[width=80mm, height=60mm]{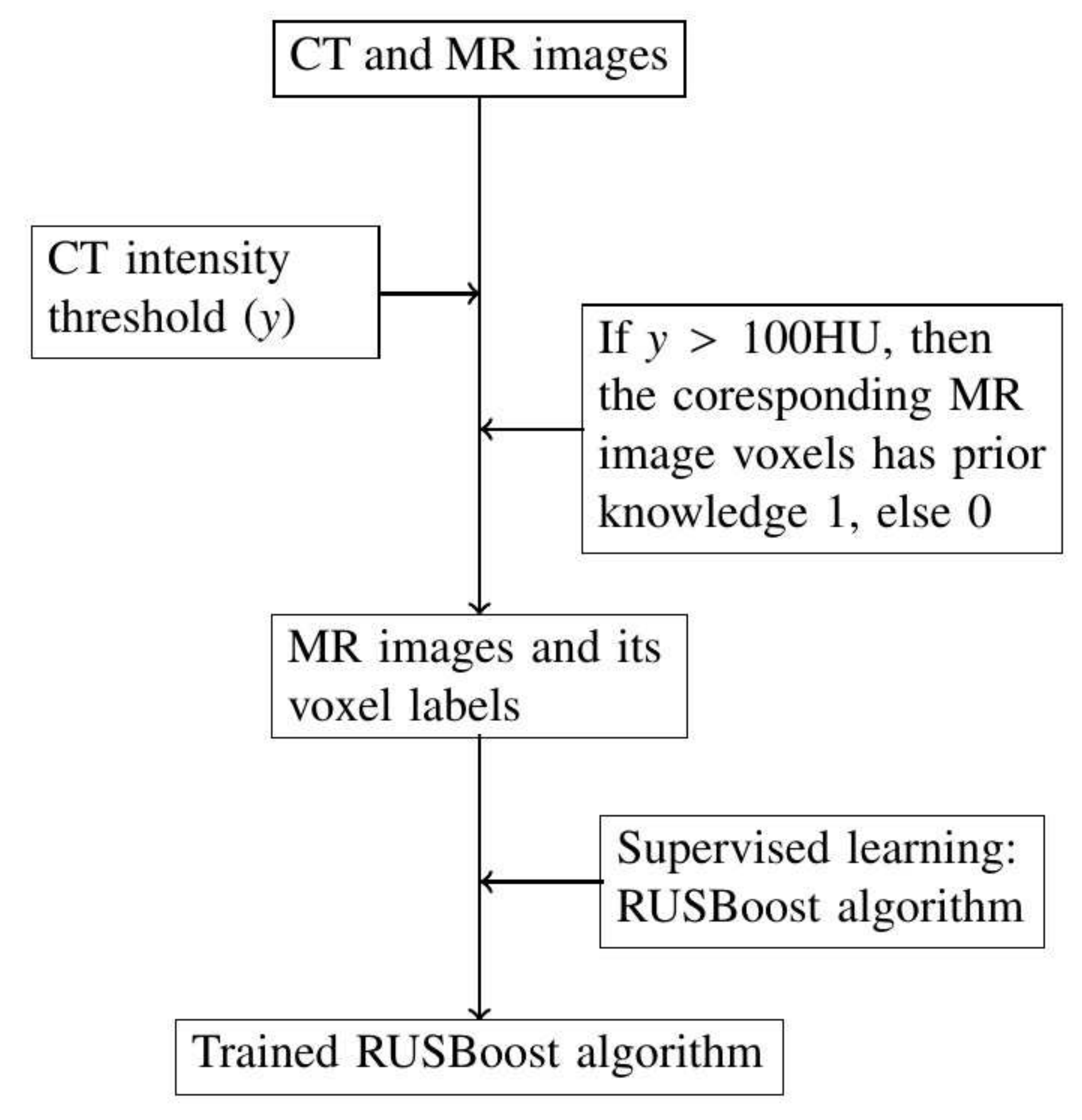}
		\caption{Conceptual model for training RUSBoost algorithm.}
		\label{fig:Pictorialvcb}
\end{figure}

\begin{figure}[H]
	\centering
	\includegraphics[width=80mm, height=60mm]{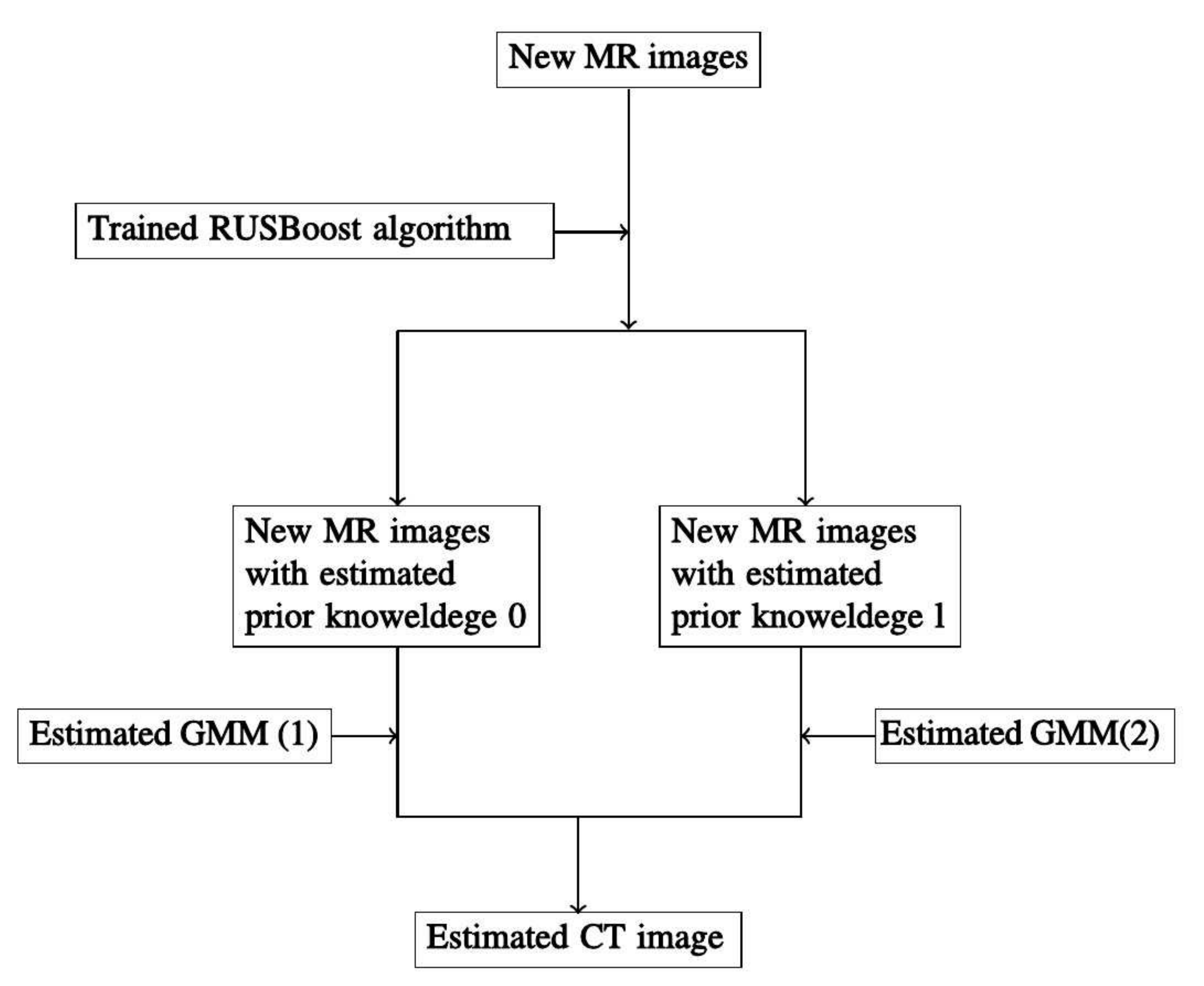}
	\caption{Conceptual model for CT image estimation.}
	\label{fig:Pictorialwsad}
\end{figure}

We summarised the RGMM in  Figs. \ref{fig:Pictorialed}, \ref{fig:Pictorialvcb}, and \ref{fig:Pictorialwsad}.  The conceptual model for RGMM involves two stages. The first stage consists of training GMM and the RUSBoost algorithm as shown in  Figs. \ref{fig:Pictorialed} and \ref{fig:Pictorialvcb}, respectively.   
The second stage is the CT image estimation, which is shown in Fig. \ref{fig:Pictorialwsad}.
\subsection{Model validation method}
To evaluate the quality of the RUSBoost classification algorithm, we use a 10-fold cross-validation method at voxel level. Using the model trained on out-of-fold observations, we can compute the classification error of the method for the in-fold observations. This is calculated by
\begin{eqnarray}
err=\frac{1}{n}\displaystyle\sum_{i=1}^{n}1_{\left\lbrace \hat{y}_{i}\ne y_{i}\right\rbrace  },
\end{eqnarray}
where
\begin{eqnarray}
1_{\left\lbrace \hat{y}_{i}\ne y_{i}\right\rbrace } =
                                   \begin{cases}
	                         	        1, & \text{if } \hat{y}_{i}\ne y_{i} ,\\
	                                 	0, & \text{otherwise},
                                	\end{cases}
\end{eqnarray}
and $\hat{y}_{i}$ is the predicted class label. The number of weak learners can be plotted against the cumulative of the averages of the classification errors in order to validate the predictive quality of the method.

An overall accuracy measure can be exploited to select the best-trained classification model. However, it is not a proper measure in imbalanced class distribution. The reason is that less prevalent classes have very little impact on the accuracy measure compared to the prevalent classes  \cite{JoshiKumarV}.  In 1994, Buckland et al. utilised precision $\left(Pr\right)$, recall $\left(Re\right)$, and F-score $\left(F\hspace{-0.03in}s\right)$ to investigate the performance of a classification method for skewed class distribution \cite{BucklandMGey}. The measures are obtained by
\begin{eqnarray}
	Pr&=&\frac{TP}{TP+FP}\quad\text{and} \quad Re=\frac{TP}{TP+FN}, \\F\hspace{-0.03in}s&=&\frac{\left( 1+\beta^2\right)\times Re\times Pr}{\beta^2\times Pr + Re},
\end{eqnarray}
where $TP$, $FP$ and $FN$ represent True Positive, False Positive, and False Negative. The positive and negative classes represent the minority and the majority classes in the data. The parameter $\beta$ denotes the relative importance of precision versus recall and is usually set to 1.  Standard classification methods usually produce classifiers that do not accurately predict the minority class. Therefore, we need to obtain a classification method that can improve the recall without affecting the precision. However, the roles of recall and precision are usually conflicting. As a result, we need to use the F-score to assess the goodness of fit of the classifier for the minority class.

We evaluate the CT image estimation performance of RGMM using a leave-one-out cross-validation method at patient level. One dataset from a patient is kept for validating the model, and the remaining datasets from the other patients are used for training the model. This procedure is repeated for each patient. For a given validation dataset, let ${Y}_{i}$; and $\hat{Y}_{i}$ be the intensity of the CT image at voxel $i$, and its estimated intensity, respectively. We choose the mean absolute error (MAE) for measuring estimation accuracy. This is a robust measure that is fairly insensitive to outliers. Better models have lower average MAEs. In addition, we use a smoothed residual plot to assess the predictive quality of the model through the tissues of the head. Smoothed residuals can be computed by moving over the CT image intensity range with non-overlapping windows, on each window we compute either the average of the residuals or their absolute values. A moving average over the non-overlapping windows is calculated to obtain the smoothed residual plot or the smoothed absolute residual plot. Smoothed residual plots enable to observe the general behaviour of the residuals and help to point out the tissues on which the statistical models are working and not working.

In summary, the MAE, smoothed residual plot, and smoothed absolute residual plot are used to compare the CT image estimation performance of RGMM with the existing model-based methods. We used the same nine datasets as in Kuljus et al. \cite{Kuljus} in order to make the comparisons possible. Kuljus et al. experienced numerical difficulties in estimating the existing model-based methods such as MRF and GMM for the nine datasets. Following that, the authors used nine and five (excluding four heads) datasets to evaluate the estimation performance of HMM, MRF, and GMM. In line with Kuljus et al., it is desirable to explore the robustness of our method based on the nine and five datasets. When comparing RGMM and HMM, all nine patients were used. When including MRF, GMM, and GMM$^{*}$ in the comparison, only five of the patients could be included due to instability in MRF and GMM. 
\section{Results}\label{sect3}
Histogram plots of the data from the nine patients showed that four of the datasets have high intensities in the MR images acquired with a 10 degree uniform flip angle. These high intensities are observed in some regions of bone tissues.
\subsection{Result: Supervised learning method}
\begin{figure}[H]
	\centering
	\includegraphics[width=80mm, height=40mm]{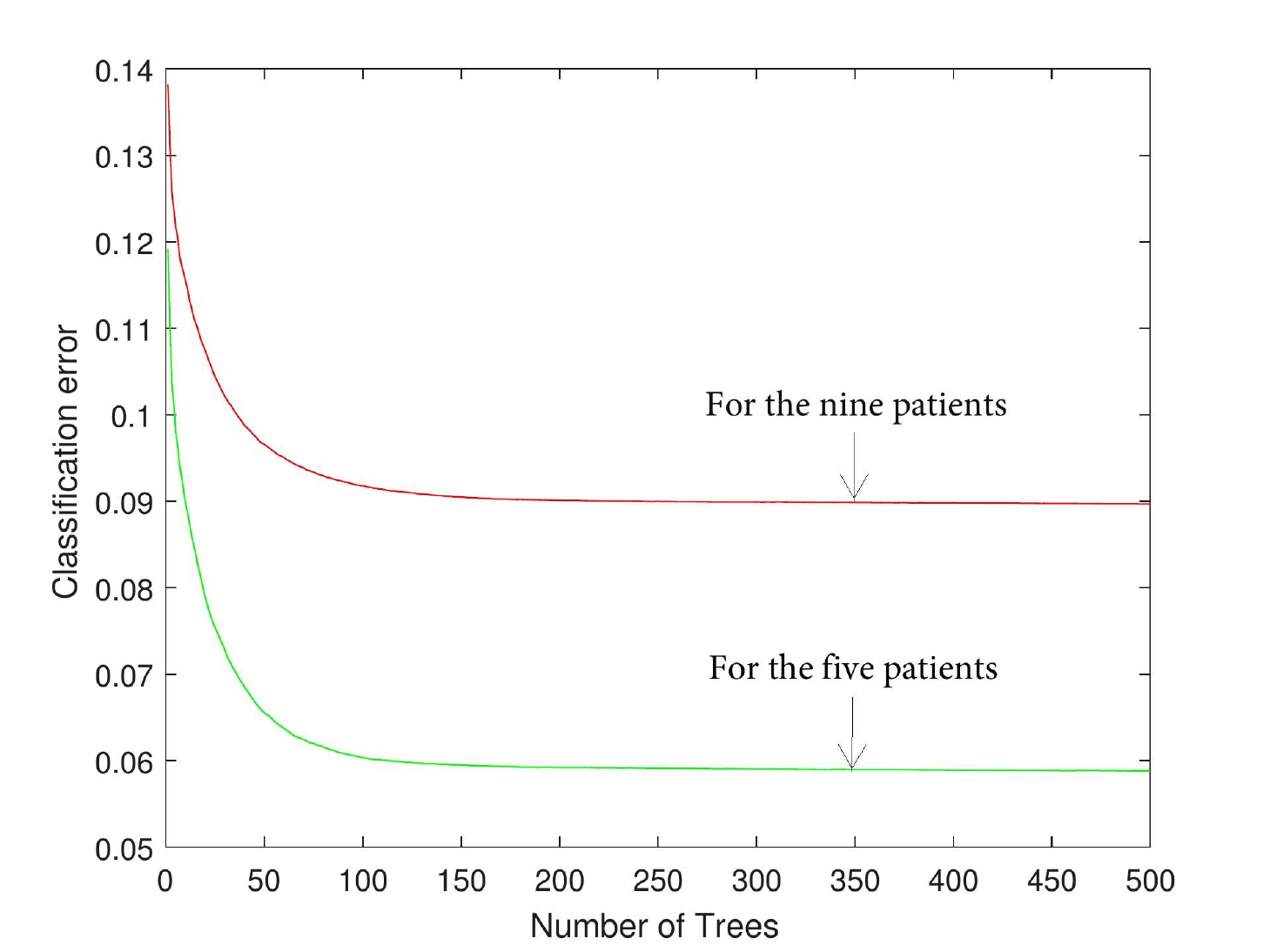}
	\small\caption{Classification error of the RUSBoost algorithm for the nine patients and the five patients.}
	\label{Or3378}
\end{figure}

We utilised CT image intensities to label each voxel of its corresponding MR images as 0 or 1. For the nine datasets, it showed that 18.49\% of instances belong to the minority class (with label 1). We began the training of the RUSBoost method with 500 decision trees and explored the training using several splits and leaf sizes for the decision trees. The investigation suggested that a maximum of 400 splits and a minimum of 5 leaf sizes were enough to obtain the desired number of decision trees. Using 10-fold cross-validation, the evaluation of the predictive quality of the algorithm revealed that it can achieve a classification error of under 9.05\% by employing 150 or more decision trees. The classification error of our method is shown in Fig. \ref{Or3378}. We used the F-score to select the best model. The predictive performance of the method depends on the number of extracted spatial features, and when we consider the six extracted spatial features, the best model has an F-score of 74.07\% with a corresponding overall accuracy of  90.05\%. On the other hand, the best model has an F-score of 76.48\% when we take the twenty-six extracted spatial features into account. In this case, the overall accuracy of the best model is 91\%. Although the change in F-score is small, the effect of using the twenty-six spatial features on CT image estimation is non-negligible.

For the five datasets, the data indicate that 17.45\% of the instances belong to the minority class. Under the same setting as for nine datasets, the method achieves a classification error of under 5.95\% by employing 150 or more trees. Fig. \ref{Or3378} shows that the best model has an F-score of 81.61\% when the six extracted spatial features are taken into account, and it has a 93.55\% overall accuracy. When we consider the twenty-six extracted spatial features, the best model has an F-score of 83.27\% and an overall accuracy of 94.12\%.
The F-score and overall accuracy appear to depend on the number of extracted spatial features. Hereinafter, we use the trained RUSBoost algorithm consisting of 150 trees and take the twenty-six extracted spatial features into account.
\subsection{Result: RGMM}
In this section, we present results obtained by RGMM and compare its performance with the existing model-based methods.

\begin{figure}[H]
	\centering
	\includegraphics[width=80mm, height=40mm]{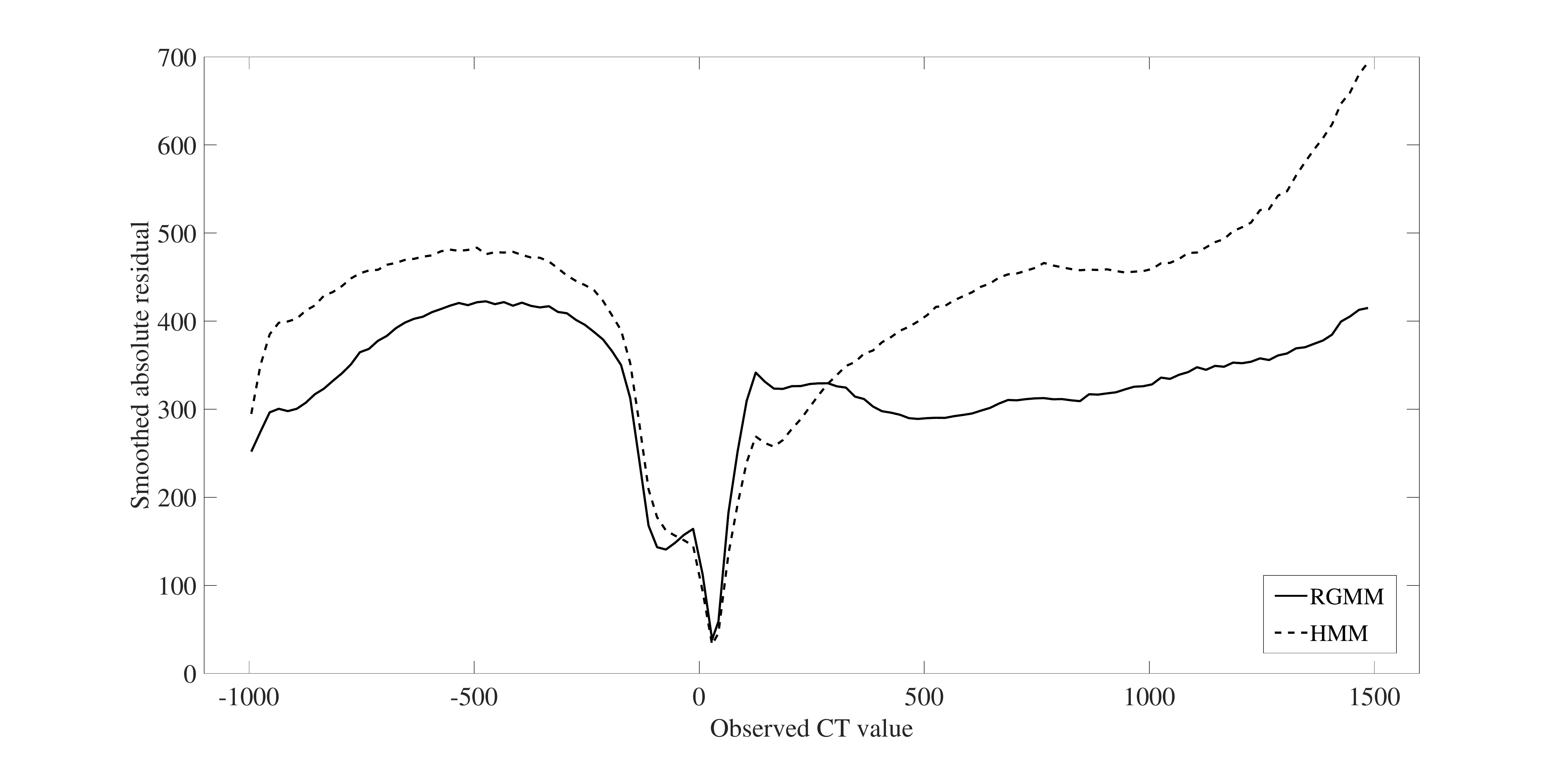}
	\caption{Smoothed absolute residual plot in HU for the nine patients.}
	\label{fig:absp9}
\end{figure}

\begin{figure}[H]
	\centering
	\includegraphics[width=80mm, height=40mm]{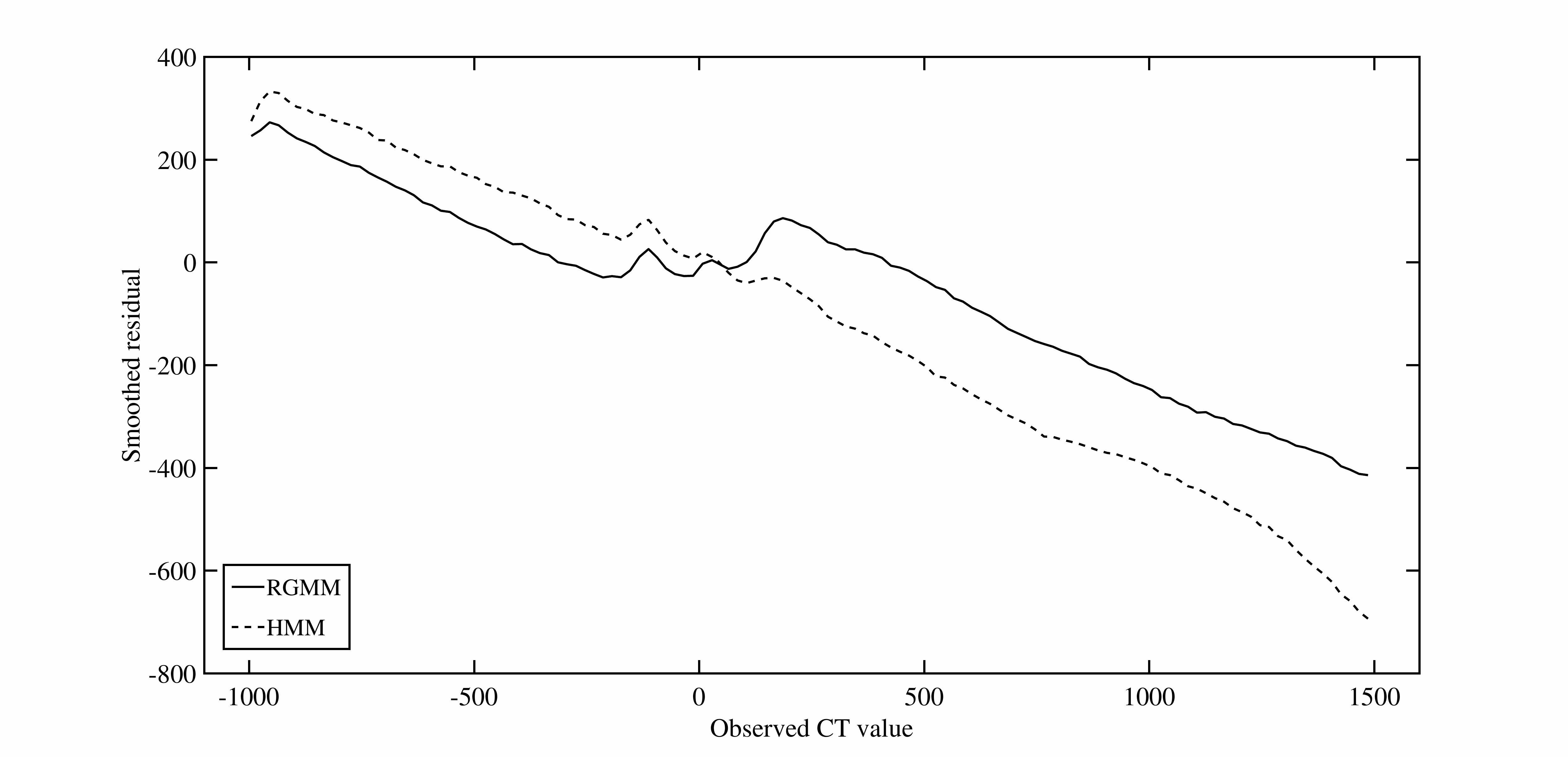}
	\caption{Smoothed residual plot in HU for the nine patients.}
	\label{fig:absp109}
\end{figure}

For the case of nine datasets, the optimal estimates of the RGMM parameters were obtained for  $J_{0}=5$ and $J_{1}=6$, which are the optimal underlying numbers of classes.  According to Kuljus et al. \cite{Kuljus}, the optimal estimates of HMM parameters were obtained for 8 underlying numbers of classes. They showed that HMM outperforms GMM, and GMM is numerically less stable on the nine datasets. Kuljus et al. \cite{Kuljus} also encountered numerical difficulty in estimating the MRF for the nine datasets. Therefore, we could only compare the CT image estimation performance of RGMM and HMM for the nine datasets. Using the leave-one-out cross-validation method, we have summarised the mean absolute errors of the CT image estimation in Table \ref{tab:Overall1}. 
\begin{table}[H]
	\centering
	\captionsetup{type=table}
	\caption{Summary of  the mean absolute errors in HU of the estimation for the nine patients.}
	\label{tab:Overall1}
	\begin{tabular}{c|ccccc}
		\hline \hline
		Patient&RGMM &HMM  \\\hline
		1& 145.80&146.31\\
		2& 139.51&146.15\\
		3& 291.76&297.35\\
		4& 143.21&157.00\\
		5& 179.50&259.67\\
		6& 152.02&199.34\\
		7& 302.08&351.73\\
		8& 140.45&153.21\\
		9& 155.49&153.87\\\hline
		mean & $\mathbf{183.31}$&$\mathbf{207.18}$\\\hline\hline
	\end{tabular}
\end{table}
Table \ref{tab:Overall1} shows that RGMM outperforms HMM by approximately 23 HU on average. This implies that the supervised learning method played a major role in improving CT image estimation. One of the main advantages of our method is that it is more robust across the patients than HMM. For instance, HMM has poor performance for patients 3, 5, 6, and 7, especially on bone tissues, see Table \ref{tab:Overall234}. We noticed that the MR images of the four heads acquired at 10 degree uniform flip angles had peculiarly high intensities in certain regions of bone tissues. This was not observed in the same regions of the remaining heads. Table \ref{tab:Overall234} shows the robustness of our method on bone tissue-dominated regions, and our approach improved the estimation by approximately 107 HU on average.  
	\begin{table}[H]
		\centering
		\captionsetup{type=table} 
		\caption{Summary of the mean absolute errors in HU for the nine patients in bone tissue-dominated regions.}
		\label{tab:Overall234}
		\begin{tabular}{c|ccccc}
			\hline \hline
			Patient&RGMM &HMM  \\\hline
			1&  284.93&331.11\\
			2& 295.42&359.93\\
			3& 564.65&615.97\\
			4& 269.80 &342.15\\
			5& 358.15&724.20\\
			6& 303.35&498.62\\
			7& 557.98&616.41\\
			8& 238.37&296.02\\
			9& 297.64&355.82\\\hline
			mean & $\mathbf{352.25}$&$\mathbf{460.03}$\\\hline\hline
		\end{tabular}
	\end{table}
We merged the CT image intensities of the nine patients and denoted this by $mCT$. Similarly, we merged the estimated CT image intensities of the nine patients and represented this by $sCT$.  Over non-overlapping windows on $mCT$ with a window size of 20 HU, we computed  the average over the windows for both $mCT$ and $|sCT-mCT|$. Plotting the averages, we obtained the smoothed absolute residual plot shown in Fig. \ref{fig:absp9}. The smoothed plot shows that RGMM has better CT image estimation quality than HMM. In particular, RGMM has superior bone tissue estimation quality.
Plotting $mCT$ against  $sCT-mCT$, we obtained the smoothed residual plot shown in Fig. \ref {fig:absp109}. Compared to HMM, our method reduced both the underestimation and overestimation problem in CT image estimation. 
To compare the robustness of our method with the existing model-based methods, it is desirable to further investigate CT image estimation on the remaining five datasets 1, 2, 4, 8, and 9. We obtained the optimal estimates of RGMM parameters for $J_{0}=6$ and $J_{1}=6$, which are the optimal underlying number of classes. According to Bayisa and Yu \cite{FekaduLB}, the optimal estimates of HMM, MRF, GMM, and GMM$^{*}$ parameters are obtained for an optimal underlying number of classes of 5 in HMM and MRF and an optimal underlying number of classes of 8 in GMM, and 6 in GMM$^{*}$ (on both bone and non-bone tissues). Using the leave-one-out cross-validation method, we have summarised the mean absolute errors of the models in Table \ref{tab:Overall2}.
\begin{table}[H]
	\vspace{-0.065in}
	\caption{Summary of  mean absolute errors in HU of the estimation for the five patients.}
	\label{tab:Overall2}
	\centering
	\begin{tabular}{{p{0.05\textwidth}|p{0.05\textwidth}p{0.05\textwidth}p{0.05\textwidth}p{0.05\textwidth}p{0.05\textwidth}}}
		\hline \hline
		Patient&RGMM  &GMM$^{*}$ &HMM  &MRF  &GMM  \\\hline
		1	& 135.72 &144.58&133.56&137.95& 145.67\\
		2	& 134.31 &152.04& 138.68 & 137.34&153.40 \\
		4	& 140.01 &161.28&142.92 & 158.99 & 163.85 \\
		8	& 125.35 &146.62& 139.17 &  145.05 &146.78\\
		9	&133.48 & 160.62&  143.89 &165.36&158.85\\\hline
		mean  &$\mathbf{133.77}$& $\mathbf{153.03}$& $\mathbf{139.65 }$ &$\mathbf{148.94}$ & $\mathbf{153.71}$\\\hline\hline
	\end{tabular}
\end{table}
Table \ref{tab:Overall2} shows that RGMM improved the estimation quality by approximately 6 HU on average and that RGMM is more stable. Most importantly, RGMM has much better performance on tissues dominated by bone (tissues that have CT image intensities greater than 100 HU).  Table \ref{tab:Bone_region1} summarises the mean absolute errors of the models for tissues dominated by bone.
\begin{table}[H]
	\caption{Mean absolute errors in HU for tissues dominated by bone.}
	\label{tab:Bone_region1}
	\hspace*{-0.2cm}
	\centering
	\begin{tabular}{p{0.05\textwidth}|p{0.05\textwidth}p{0.05\textwidth}p{0.05\textwidth}p{0.05\textwidth}p{0.05\textwidth}}
		\hline \hline
		Patient&RGMM &GMM$^{*}$ &HMM  &MRF  &GMM  \\\hline
		1	&265.49 &315.16& 324.94& 307.95& 314.41 \\
		2   &273.84 & 348.05&  360.03&  328.89& 365.12\\
		4   &236.20 & 301.78&  331.16&  322.48&  328.01 \\
		8	&216.67&269.33&  296.04&  280.63&  292.78\\
		9	&282.27&349.28&  366.13&  357.22&  359.56\\\hline
		mean&$\mathbf{254.89}$& $\mathbf{316.72}$& $\mathbf{335.66}$ &$\mathbf{319.43}$ &$\mathbf{331.98}$\\\hline\hline
	\end{tabular}
\end{table}

\begin{figure}[H]
	\centering
	\includegraphics[width=80mm, height=40mm]{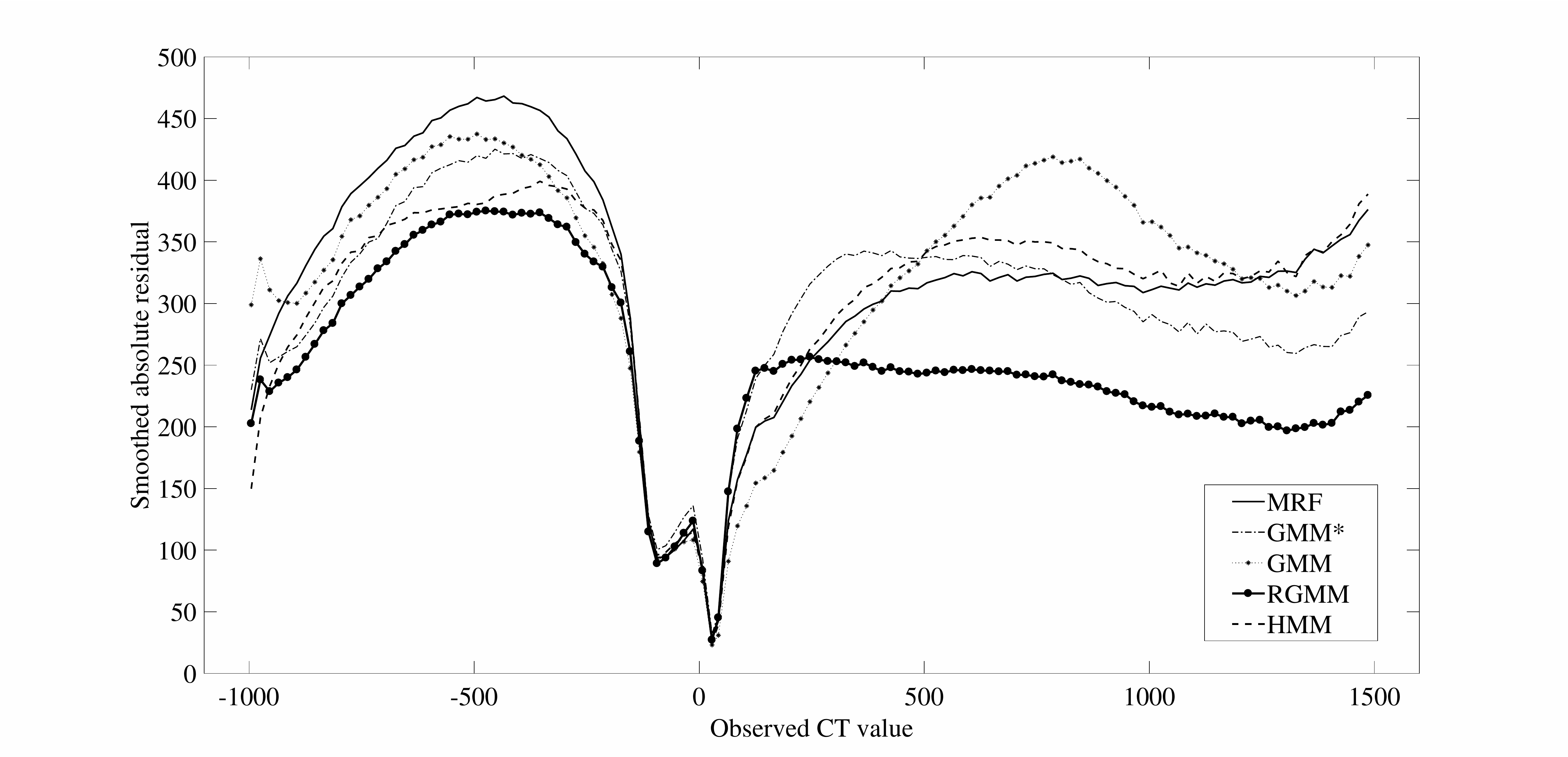}
	\small\caption{Smoothed absolute residual plot in HU for the five patents.}
	\label{Or3}
\end{figure}

\begin{figure}[H]
	\centering
	\includegraphics[width=80mm,height=40mm]{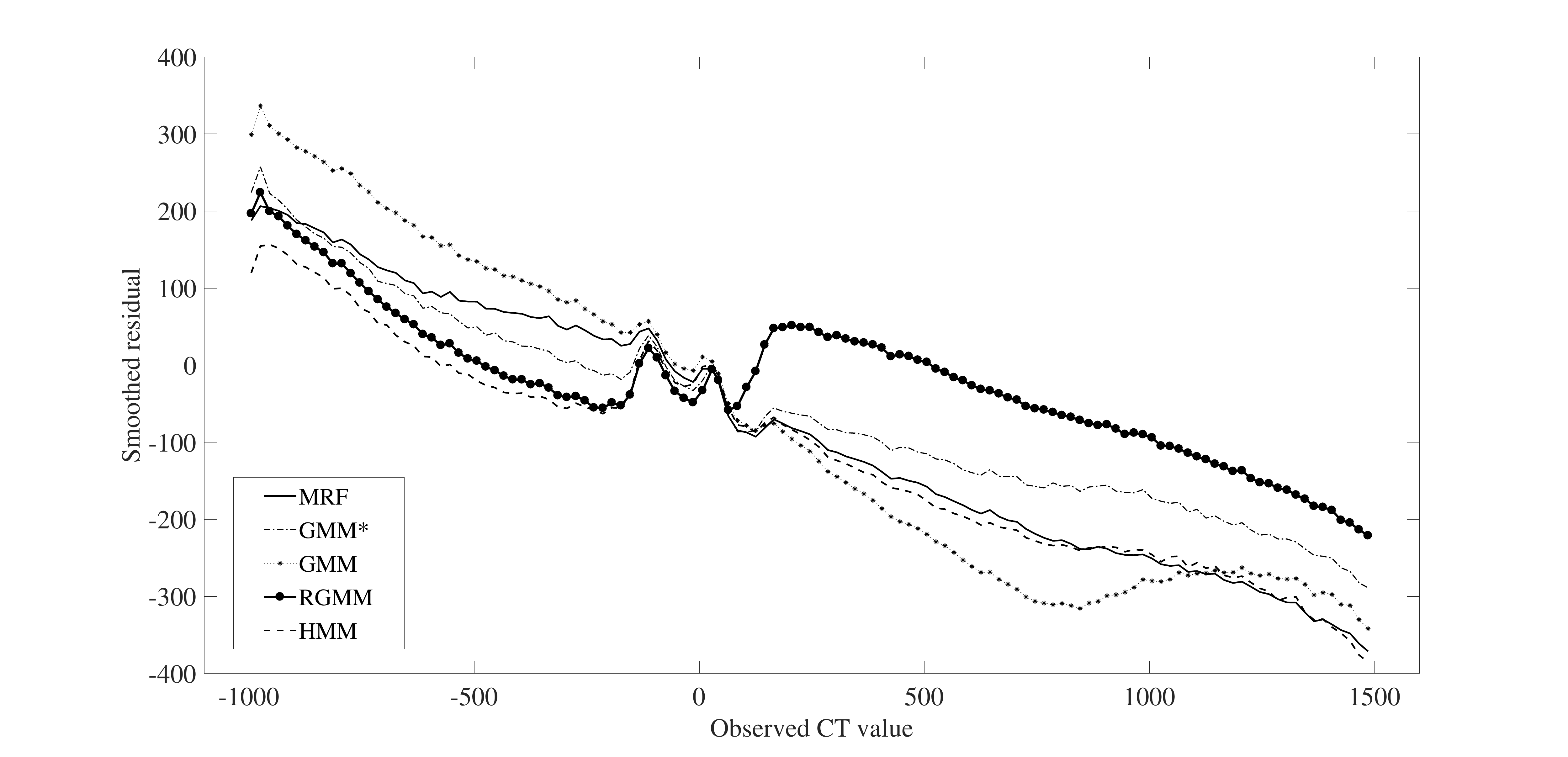}
	\small\caption{Smoothed residual plot in HU for the five patients.}
	\label{Or4}
\end{figure}

Table \ref{tab:Bone_region1} shows that RGMM improved the estimation quality in bone tissue-dominated regions by approximately 62 HU on average.
We also utilised a moving average to evaluate the performance of the models on the tissues of the heads. Fig. \ref{Or3} shows the smoothed absolute residual plot for the five patients that was obtained in a similar manner as the smoothed plot for nine patients.
The smoothed plot shows that RGMM has better performance than the other models. Specifically, our method has better bone tissue estimation quality. We also used the smoothed residual plot shown in Fig. \ref{Or4} to investigate the nature of the models for the tissues. 
The residual plot demonstrates that the models underestimated for the bone tissues and overestimated for the air spaces of the heads. However, it is evident from the plot that RGMM reduced the underestimation problem for bone tissues compared to the other methods used in this work.

\begin{figure}[H]
	\centering
	\hspace*{-0.8cm}
	\includegraphics[width=80mm, height=40mm]{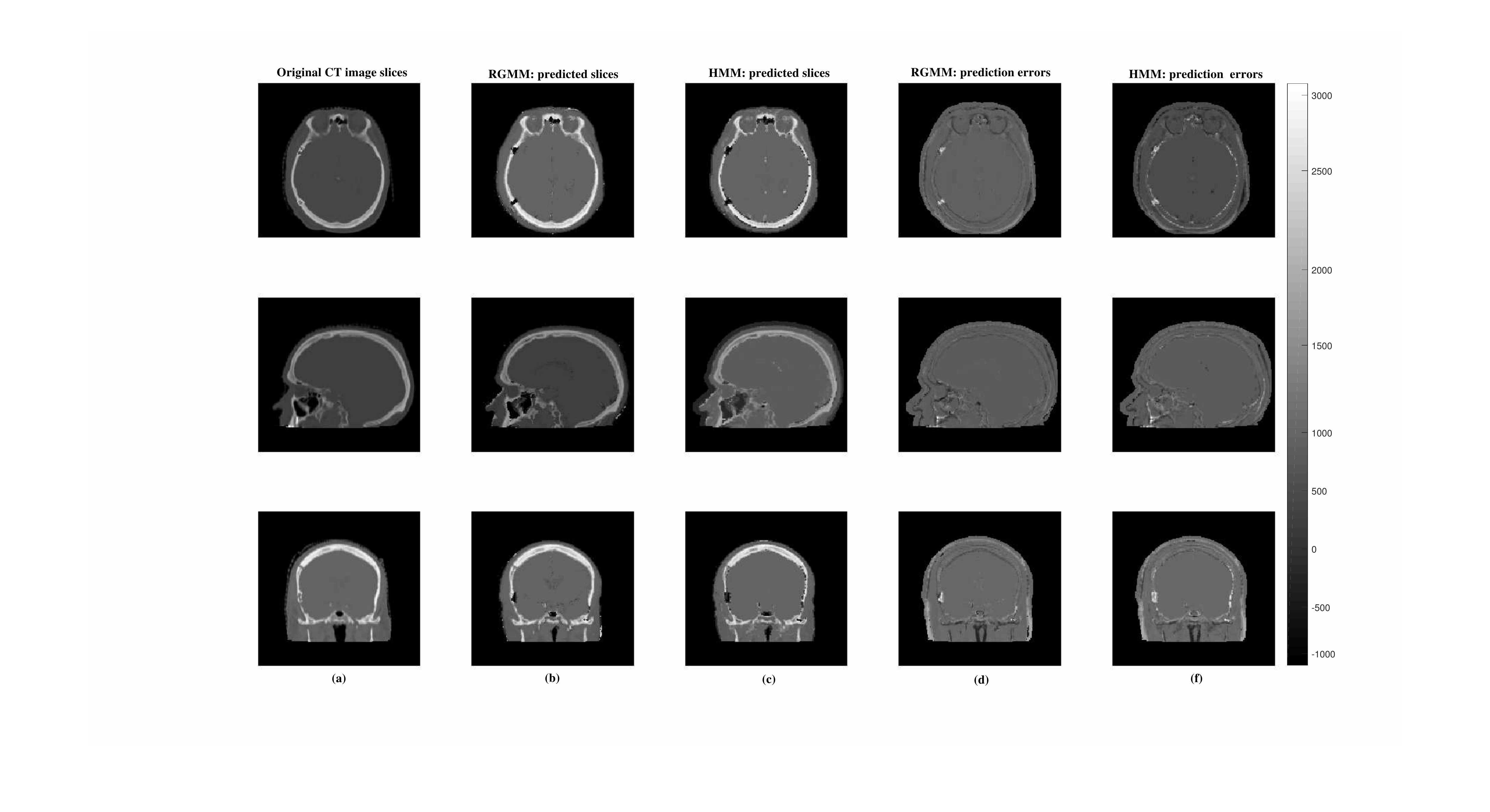}
	\vspace*{-0.5cm}
	\small\caption{The first column (a) represents slices of CT image, the second and third column (b) and (c) shows the corresponding  predicted slices of CT images, and the last two columns (d) and (e) show the errors of the prediction.}
	\label{Or45}
\end{figure}

Existing methods have poor performance in the region of the throat cavity \cite{Johansson2, HildemanA}, and we observed similar behaviour in our method. The results of our approach and HMM are presented in  Fig. \ref{Or45} showing slices of a CT image, its predicted images, and the associated prediction errors, which are conventionally defined as $mCT-sCT$, for a representative patient.
\section{Discussion} \label{sec4}
We have presented a statistical learning method for CT image estimation from MR images and have evaluated the method using cross-validation on nine and five datasets.

Several methods have previously been utilised to investigate CT image estimations \cite{Adam, Nie, Huynh,Arabi,Kuljus, FekaduLB}. Our method is different from these works in that it combines supervised statistical learning and GMM to generate substitute CT images. Our approach also differs from the studies carried out by Keereman et al. \cite{KeeremanV}, Catana et al. \cite{CatanaCcvf}, and Hofmann et al. \cite{HofmannMo} because we used UTE MRI sequences with two flip angles at two different echo times and used RGMM to estimate the CT images. 

Keereman et al. \cite{KeeremanV} suggested that visualization of cortical bone with MRI is possible with UTE sequences, and their results revealed that UTE approach can also enable the  separation of bone and air, which is problematic with traditional MR techniques. Even though bones are imaged with weak intensity and poor resolution using UTE sequences, the use of UTE sequences in MRI is advantageous for our method as it makes the statistical modeling feasible for CT image estimations.  According to Johansson et al. \cite{johansson2012voxel}, the use of UTE sequences with two flip angles can substantially increase the quality of the CT image estimations. However, streak artifacts in the images from the UTE sequences become more severe when larger body parts are imaged \cite{Johansson2}. Even though the UTE pulse sequences can be simple in theory, its successful implementation needs accurate timing and a detailed understanding of the hardware performance  \cite{margosian2012practical}.

The proposed method has resulted in better accuracy of CT image estimation, especially for bone tissues, which is the most difficult task in the estimation of CT images from MR images  \cite{Karlssonr}. Figs. \ref{fig:absp9} and \ref{Or3} and Table \ref{tab:Overall1}--\ref{tab:Bone_region1} show that our method has better CT image estimation quality compared to the existing model-based methods. We observed that the absolute errors in CT image estimation are not only severe in the bone and air regions, but also at their interfaces with soft tissues. This problem might not be directly attributed to the model and might instead be related to patient motion during data acquisition, changes in the anatomy of the patients, and uncertainties during image registration. In other words, the model has difficulty to handle the sudden changes of tissue-interfaces. In order to take this into account, more data are required for the model to capture the anatomical changes. Regardless of this issue,  our approach outperforms the already established results. By reducing the aforementioned effects, one can even increase the performance of our method.

Our method is more stable than the existing model-based methods. Table \ref{tab:Overall1} shows that the variation in the mean absolute errors of the estimation across the heads for RGMM is lower than for HMM, and our method improved CT image estimation by approximately 23 HU on average. In particular, it provided better quality in estimating bone tissue-dominated regions with an improvement by approximately 107 HU on average (Table \ref{tab:Overall234}). In other words, it has improved CT image estimation by $5\%$ and $23\%$ on whole-brain and bone tissues respectively. For the nine datasets, the estimations by GMM and MRF are unstable due to ill- conditioning and the inverse problem of covariance matrices in these models \cite{Kuljus}. This is also the reason that we did not compare our method with GMM$^{*}$, that is, we did not expect good partitioning of the datasets based on the estimated CT image by GMM on the nine datasets. Furthermore, Table \ref{tab:Overall2} shows that our method is better than GMM$^{*}$, and it is reasonable to expect a similar result on the nine datasets. RGMM is less sensitive than HMM, MRF, and GMM on heads 3, 5, 6, and 7. The reason for the sensitivity of HMM, MRF, and GMM for the four datasets is that the MR images of the four heads obtained at 10 degree uniform flip angles have strange proton densities in certain regions of bone tissues. 
 
To further investigate the robustness of our method, we compared it with the existing model-based methods and explored the estimation of CT images using the remaining five subset datasets.  Tables \ref{tab:Overall2}--\ref{tab:Bone_region1} show that our method is more stable than the other models. Moreover, it has improved the estimation of bone tissue-dominated regions by approximately 62 HU on average. This shows that our method is relatively consistent and is a promising approach for generating substitute CT images from MR images.

In general, distortion in the MR image volumes in addition to motion could affect the accuracy of CT estimation. Since this impact will be more or less valid for all methods, the relative performance would keep similar. The UTE images also have low distortion due to the short echo time, and the gradient non-linearity correction may have some residual distortion but it should be small. Moreover, our method is more robust which would imply a less impact. 

Johansson et al. \cite{Johansson2} used UTE MR images. We also used UTE MR images in this study. Recently, zero-echo-time (ZTE) has been shown to provide a sequence of MR images that can exhibit sufficient contrast between different tissues, especially air and bone. These sequences can be employed to obtain better estimations of CT images.

An interesting extension of our work is to relax the boundary of the class labelling and to instead try to learn the boundary from the datasets. Another alternative to extend this work is to use supervised learning at the two stages that were utilised to estimate CT images from MR images. Furthermore, although the error metrics used in this study, e.g. MAE, are easy to understand and consistent with Johansson et al. \cite{Johansson2}, it is by no means the endpoints for evaluating the quality of substituted CT. Thus, it is certainly necessary to conduct a further investigation to quantify the quality of substituted CT from the clinical point of view, such as dose calculation accuracy and accuracy of attenuation correction.
 
\section{Conclusions}\label{sec5}
We have shown that our method has better CT image prediction quality, especially in bone tissue estimation, compared to previous model-based methods. Evaluation of our method shows that it predicts CT image intensities accurately and that it is a promising method to obtain CT image substitutes for the implementation of fully MRI-based radiotherapy and for PET/MRI applications.

 \section*{Acknowledgments}
\noindent We would like to thank two anonymous referees, the Associate Editor, and the Editor for their detailed and insightful comments and suggestions that help to improve the quality of this paper.  
This work is supported by the Swedish Research Council grant (Reg. No. 340-2013-5342). The authors thank Adam Johansson and Thomas Asklund for providing us data and David Bolin for providing us  Mathlab code for MRF. Moreover, the authors thank Kristi Kuljus for HMM results.
The computations were performed on resources provided by the Swedish National Infrastructure for Computing (SNIC) at High Performance Computing Center North (HPC2N).
 \section*{Disclosure of Conflicts of Interest}
 The authors have no relevant conflicts of interest to disclose.
\section*{References}
\bibliographystyle{plainnat}
\bibliography{sample_3}
\end{document}